\documentclass[prb,floatfix,twocolumn,showpacs,aps]{revtex4}
\usepackage{graphicx}
\usepackage{dcolumn}
\usepackage{amsmath}
\usepackage{natbib}
\usepackage{amsfonts}
\usepackage{amsmath}
\usepackage{amssymb}
\usepackage{graphicx}

\begin{document}

\title{Heat capacity studies of Ce and Rh site substitution in the heavy
fermion antiferromagnet CeRhIn$_{5}$: Short-range magnetic interactions and
non-Fermi-liquid behavior}
\author{B.E. Light, Ravhi S. Kumar and A.L. Cornelius}
\affiliation{Department of Physics, University of Nevada, Las Vegas, Nevada, 89154-4002}
\author{P.G. Pagliuso$^{\ast }$ and J.L. Sarrao}
\affiliation{Materials Science and Technology Division, Los Alamos National Laboratory,
Los Alamos, NM\ 87545}

\begin{abstract}
In heavy fermion materials superconductivity tends to appear when long range
magnetic order is suppressed by chemical doping or applying pressure. Here
we report heat capacity measurements on diluted alloyes of the heavy fermion
superconductor CeRhIn$_{5}$. Heat capacity measurements have been performed
on CeRh$_{1-y}$Ir$_{y}$In$_{5}$ ( $y\leq 0.10$) and Ce$_{1-x}$La$_{x}$RhIn$%
_{5}$ ($x\leq 0.50$) in applied fields up to 90 kOe to study the affect of
doping and magnetic field on the magnetic ground state. The magnetic phase
diagram of CeRh$_{0.9}$Ir$_{0.1}$In$_{5}$ is consistent with the magnetic
structure of CeRhIn$_{5}$ being unchanged by Ir doping. Doping of Ir in
small concentrations is shown to slightly increase the antiferromagnetic
transition temperature $T_{N}$ ($T_{N}=3.8$ K in the undoped sample). La
doping which causes disorder on the Ce sublattice is shown to lower $T_{N}\,$%
\ with no long range order observed above 0.34 K for Ce$_{0.50}$La$_{0.50}$%
RhIn$_{5}$. Measurements on Ce$_{0.50}$La$_{0.50}$RhIn$_{5}$ show a
coexistence of short range magnetic order and non-Fermi-liquid behavior.
This dual nature of the Ce $4f$-electrons is very similar to the observed
results on CeRhIn$_{5}$ when long range magnetic order is suppressed at high
pressure.
\end{abstract}

\date{\today}
\maketitle

\section{Introduction}

Among heavy fermion (HF) materials, magnetically mediated superconductivity
has been observed in many materials at the point where long range order is
suppressed by alloying or applying pressure at a quantum critical point
(QCP).\cite{Steglich79,Jaccard92,Movshovich96,Grosche96,Mathur98,Hegger00}
One of these systems, CeRhIn$_{5}$, is antiferromagnetic (AF)\ at ambient
pressure with $T_{N}=3.8$ K and $\gamma \thickapprox 400$ mJ/mol K$^{2}$.%
\cite{Hegger00,Cornelius00} The AF\ state is suppressed at a pressure of
around 1.2 GPa and coexists over a limited pressure range with the
superconducting (SC) state.\cite{Hegger00,Fisher02,Mito01,Mito03} Recently,
HF systems with the formula Ce$M$In$_{5}$ ($M=$ Co and Ir) have also been
reported to become superconductors at ambient pressure.\cite%
{Petrovic01,Petrovic01_2} Unlike most HF\ superconductors, the system CeRh$%
_{1-y}$Ir$_{y}$In$_{5}$ displays a coexistence of AF\ order and SC state
over a wide range of doping $(0.3<x<0.6)$.\cite{Pagliuso01} Thermodynamic,%
\cite{Cornelius00} NQR,\cite{Curro00} and neutron scattering \cite%
{Bao00,Bao03_2} experiments all show that the electronic and magnetic
properties of CeRhIn$_{5}$ are anisotropic in nature.

To better understand the magnetic ground state out of which SC evolves, we
have performed heat capacity measurements on both the CeRh$_{1-y}$Ir$_{y}$In$%
_{5}$ $(y\leq 0.10)$ and Ce$_{1-x}$La$_{x}$RhIn$_{5}$ $(x\leq 0.50)$
systems. The magnetic field studies for various dopings are an extension of
our previous work.\cite{Cornelius00,Pagliuso02_2} The measurements were
performed along both the tetragonal $a$ and $c$ axes in applied magnetic
fields to 90 kOe. The dependence of the magnetic transitions with respect to
temperature, magnetic field applied along different crystal directions, and
doping using heat capacity measurements allows the determination of the
magnetic interactions in these complicated materials. Precisely determining
the changes of magnetic properties with different variables gives insight
into favorable conditions for magnetically mediated superconductivity. Field
induced transitions when the magnetic field is applied along the $a$
direction are seen in all samples that display AF\ order. A detailed phase
diagram for CeRh$_{0.9}$Ir$_{0.1}$In$_{5}$ shows excellent agreement to that
of the undoped parent compound CeRhIn$_{5}$ suggesting that Ir doping does
not change the AF\ order from the measured incommensurate spin density wave.%
\cite{Curro00,Bao00,Bao03_2} La doping suppresses magnetic order with the $%
x=0.50$ sample showing no long range AF\ order; however, a coexistence of
short range magnetic order and non-Fermi-Liquid (NFL) behavior is observed.

Though a great deal of both experimental and theoretical work have been
performed on heavy fermion systems, a general understanding of the crossover
from the single impurity to the lattice limits has been elusive. Recently,
the observance of multiple energy scales \cite{Cornelius02_3,Ebihara03} and
the coexistence of localized and delocalized $f$-electrons \cite%
{Nakatsuji02,Nakatsuji03} have shown the necessity of including the lattice
and considering the localization of $f$-electrons in heavy fermion systems.
Our results on Ce$_{0.50}$La$_{0.50}$RhIn$_{5},$ which is near the QCP, show
signatures of the coexistence of short range (short range magnetic order)
and long range (non-Fermi liquid) behavior. We find a striking resemblance
of the results on Ce$_{0.50}$La$_{0.50}$RhIn$_{5}$ to those on CeRhIn$_{5}$
driven to a QCP\ under pressure.\cite{Fisher02}

\section{Experiment}

CeRh$_{1-y}$Ir$_{y}$In$_{5}$ and Ce$_{1-x}$La$_{x}$RhIn$_{5}$ single
crystals were grown by a self flux technique.\cite{Moshopoulou01} The
samples were found to crystallize in the primitive tetragonal HoCoGa$_{5}$%
-type structure \cite{Grin79,Grin86} with lattice parameters, determined by
x-ray diffraction, in agreement with literature values.\cite%
{Pagliuso01,Moshopoulou01} Heat capacity measurements, using a standard
thermal relaxation method, were performed in a Quantum Design PPMS\ system
equipped with a superconducting magnet capable of generating a 90 kOe
magnetic field. The lattice heat capacity was determined by measuring LaRhIn$%
_{5}$ which has no $f$-electrons. The LaRhIn$_{5}$ data was subtracted from
Ce$_{1-x}$La$_{x}$Rh$_{1-y}$Ir$_{y}$In$_{5}$ to obtain the magnetic heat
capacity $C_{m}$. This makes the assumption that the specific heat of the
lattice is unchanged by the substitutions. Since the lattice constants
change less than $0.6\%$ for the studied samples \cite%
{Pagliuso01,Pagliuso02_2} one expects that the Debye temperature, which is
known to depend on volume, and hence the lattice contribution to the heat
capacity remains constant for our purposes.

\section{Results and Discussion}

\subsection{Low Temperature Specific Heat}

The low temperature specific heat measurements were performed over the
temperature range 0.34 K$<T<20$ K in applied magnetic fields to 90 kOe. As
previously mentioned, the lattice contribution to the heat capacity is
subtracted using LaRhIn$_{5}$ as a reference compound. The total magnetic
specific heat can be written as 
\begin{equation}
C_{m}=C_{elec}+C_{order}+C_{hyp}  \label{cp eq}
\end{equation}%
where $C_{elec}$ is the electronic contribution, $C_{order}$ is from
magnetic correlations (short and long ranged) between the Ce $4f$-electrons,
and $C_{hyp}$ is from the nuclear moment of the In atoms. The electronic
contribution is given by $\gamma T$ for $T>T_{N}$ and $\gamma _{0}T$ for $%
T<T_{N}$ , where $\gamma >\gamma _{0}.$ In the magnetically ordered samples
below $T_{N}$, as done before,\cite{Cornelius00} we use the form%
\begin{equation}
C_{order}=\beta _{M}T^{3}+\beta _{M}^{\prime }\left(
e^{-E_{g}/k_{B}T}\right) T^{3}  \label{magnon}
\end{equation}%
where $\beta _{M}T^{3}$ is the standard AF magnon term and the second term
is an activated AF\ magnon term. The need for an activated term to describe
heat capacity data has been seen before in other Ce and U compounds \cite%
{Cornelius00,Bredl87,Dijk97,Murayama97}, and rises from an AF SDW with a gap
in the excitation spectrum due to anisotropy. The CeRhIn$_{5}$ magnetic
structure indeed displays an anisotropic SDW with modulation vector
(1/2,1/3,0.297) \cite{Bao00,Bao03_2} which is consistent with this picture.
The In atoms have a nuclear magnetic moment which gives rise to hyperfine
contribution to the heat capacity $C_{hyp}$. $C_{hyp}$ is given by $A/T^{2}$
with $A$ given by the relation \cite{Lees99}%
\begin{equation}
A=\frac{R}{3}\left(  \frac{I+1}{I}\right)  \left(  \frac{\mu H_{hyp}}{k_{B}%
}\right)  ^{2},\label{chyp}%
\end{equation}
where $I$ is the nuclear moment (9/2 for In), $\mu $ is the nuclear magnetic
moment (5.54 $\mu _{N}$ for In), and $H_{hyp}$ is the magnitude of the
internal field strength at the In site that can be due to both internal $%
H_{int}$ and externally applied $H$ fields.

\begin{figure}[tbp]
\includegraphics[width=2.7in]{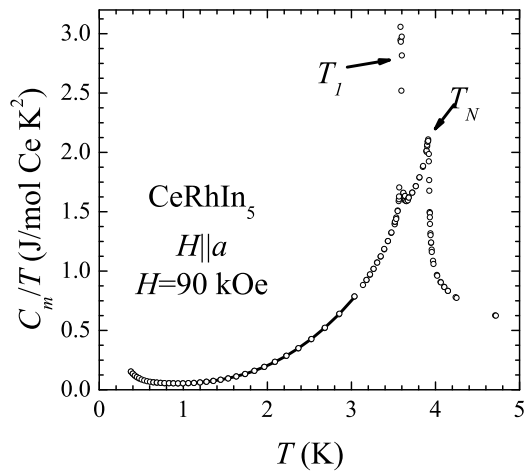}
\caption{Magnetic heat capacity $C_{m}$ divided by temperature $T$ versus $T$
measured on CeRhIn$_{5}$ in an applied magnetic field of 90 kOe with $H||a$.
Two phase transitions $T_{N}$ and $T_{1}$ correspond to the
antiferromagnetic ordering temperature and a field induced transition. The
solid line for $T<0.8$ $T_{N}$ represents a fit to the data as described in
the text.}
\label{cmfit}
\end{figure}

Data for CeRhIn$_{5}$ in a 90 kOe magnetic field applied along the
tetragonal $a$ axis ($H||a$) is shown in Fig. \ref{cmfit}.

Two phase transitions at $T_{N}=3.91$ K and $T_{1}=3.09$ K are clearly seen. 
$T_{N}$ is the transition to long range AF\ order, and $T_{1}$ is a first
order field-induced magnetic phase transition.\cite{firstord} Field induced
transitions have been observed before in CeRhIn$_{5}$,\cite{Cornelius01} and
this topic will be discussed in detail later. The upturn at low temperatures
is due to the In nuclear Schottky term. Note that we are in the high
temperature limit for the nuclear Schottky term and use the high temperature
approximation (Eq. \ref{chyp}) that the nuclear heat capacity falls off as $%
T^{-2}$. The solid line in Fig. \ref{cmfit} is a fit to Eq. \ref{cp eq} with 
$\gamma _{0}=38\pm 2$ mJ/mol Ce K$^{2}$, $\beta _{M}=6.3\pm 0.2$ mJ/mol Ce K$%
^{4}$, $\beta _{M}^{\prime }=310\pm 20$ mJ/mol Ce K$^{4}$, $%
E_{g}/k_{B}=4.4\pm 0.2$ K. Fits like this were successful for all CeRh$%
_{1-y} $Ir$_{y}$In$_{5}$ samples in all applied fields directed along either
the $a- $ or $c-$axis. However, fits to the Ce$_{1-x}$La$_{x}$RhIn$_{5}$
data, at least for $x>0.03$, did not give satisfactory results due to short
range order and non-Fermi liquid effects and only the $x=0$ and $x=0.03$
data were fit and will be reported.

\subsection{Magnetic Entropy}

\begin{figure}[tbp]
\includegraphics[width=2.7in]{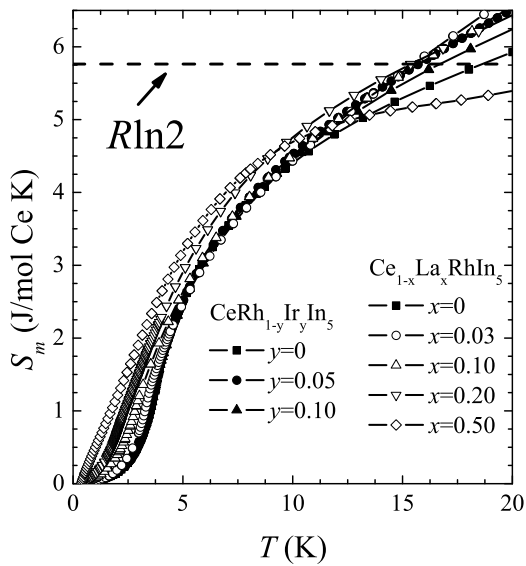}
\caption{Magnetic entropy $S_{m}$, found by integrating $C_{m}/T$, measured
on doped CeRhIn$_{5}$ samples. All of the samples approach $R\ln 2$ of
entropy by 20 K consistent with a $S=1/2$ doublet crystal-field ground
state. }
\label{entropyall}
\end{figure}

The magnetic entropy $S_{m}$ can be found by integrating $C_{m}/T$ as a
function of temperature. This has been done for all of the measured samples
and the results in zero applied field are shown in Fig. \ref{entropyall}.

All of the measured samples approach $R\ln 2$ by 20 K with the exception of
the Ce$_{0.50}$La$_{0.50}$RhIn$_{5}$ sample (which is still very close to $%
R\ln 2 $). Application of a magnetic field has a nearly negligible effect on
the measured entropy for all samples with again the exception of the Ce$%
_{0.50}$La$_{0.50}$RhIn$_{5}$ sample. The reason for the differences in the
behavior of the Ce$_{0.50}$La$_{0.50}$RhIn$_{5}$ sample relative to the
others will be explained in detail later but is due, at least in part, to
non negligible entropy below 0.35 K that we cannot measure. It is also
possible that the stoichiometry is slightly less than the nominal starting
value of $x=0.50$. In fact a value of $x=0.47$ gives an entropy of $R\ln 2$
at 20 K. For the rest of the manuscript, we will assume that $x=0.50$.
However, using $x=0.47$ has little effect on the data and would not change
our conclusions. The observance of $R\ln 2$ entropy in all of the
measurements is indicative that the measured $C_{m}$ values are due solely
to a doublet crystal-field (CF) ground state. This is consistent with other
studies which show the lowest CF\ level is a doublet separated by 60-80 K\
from the first excited level.\cite{Takeuchi01,Pagliuso02_3,Christianson02}
Thus we are confident that our measurement of $C_{m}$ are solely due to the
Ce $4f$ electrons in a $S=1/2$ doublet CF ground state.

\subsection{Magnetic Order}

\begin{figure}[tbp]
\includegraphics[width=2.7in]{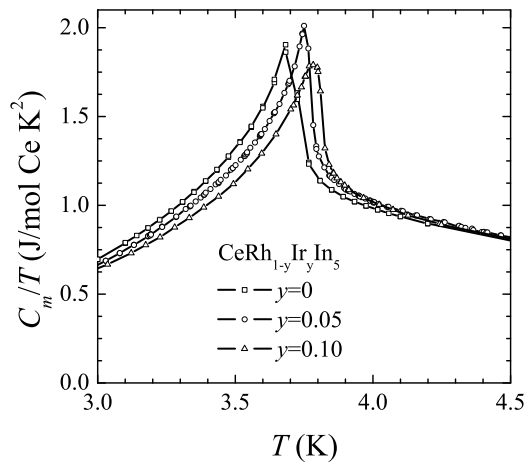}
\caption{Magnetic heat capacity $C_{m}$ divided by temperature $T$ versus $T$
measured on CeRh$_{1-y}$Ir$_{y}$In$_{5}$ in zero field. The value of $T_{N}$
increases as $y$ increases.}
\label{ceir0T}
\end{figure}

Neutron diffraction studies have shown that the substitution of 10\% La for
Ce lowers $T_{N}$ from 3.8 K found in CeRhIn$_{5}$ to 2.7 K but does not
change the SDW\ ground state.\cite{Bao02} As mentioned, CeRh$_{1-y}$Ir$_{y}$%
In$_{5}$ displays a coexistence of AF\ order and SC state over a wide range
of doping $(0.3<y<0.6)$.\cite{Pagliuso01} As it has been established that 2D
magnetic ground states favor SC,\cite{Monthoux01} it is important to
determine if the SC\ state arises out of the known SDW\ ground state of the
undoped sample. Fig. \ref{ceir0T}

shows the zero field data as a function of Ir doping.

The application of a magnetic field alters the magnetic interactions. As
reported before for CeRhIn$_{5}$,\cite{Cornelius01} when $H||c$, $T_{N}$
decreases for all of the samples that show AF\ order as is usually seen in
heavy fermion systems.\cite{Stewart84} As shown in Fig. \ref{cmfit}, for $%
H||a$ field-induced magnetic transitions are observed in CeRh$_{0.9}$Ir$%
_{0.1}$In$_{5}$. A cumulative $H-T$ phase diagram is shown in Fig. \ref%
{ceirphase}.

\begin{figure}[tbp]
\includegraphics[width=2.7in]{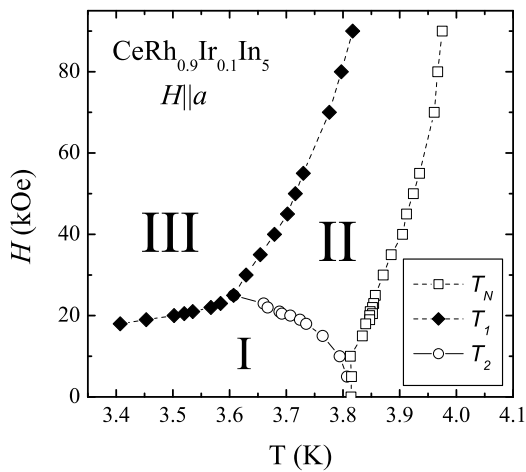}
\caption{The cumulative phase diagrams for CeRh$_{0.9}$Ir$_{0.1}$In$_{5}$ in
various applied fields $H$ applied along the $a$-axis. $T_{N}$ corresponds
to the antiferromagnetic ordering temperature, and $T_{1}$ and $T_{2}$
correspond to field-induced first- and second-order transitions
respectively. The dashed lines are guides to the eyes.}
\label{ceirphase}
\end{figure}

$T_{N}$ corresponds to the antiferromagnetic ordering temperature, and $%
T_{1} $ and $T_{2}$ correspond to field-induced first- and second-order
transitions respectively. The dashed lines are guides to the eyes. The
similarity to the phase diagram of the undoped CeRhIn$_{5}$ is remarkable.%
\cite{Cornelius01} This naturally leads to the conclusion that the Ir
substitution does not change the magnetic structure (incommensurate SDW) of
CeRhIn$_{5}$ at least for $x\leq 0.10$. In a manner similar to the effect of
doping, recent neutron scattering results show that the incommensurate SDW\
only weakly changes with pressure up to 2.3 GPa.\cite{Llobet03} As stated
previously, the magnetic structure in Regions I and II is a spin density
wave that is incommensurate with the lattice where Region II has a larger
magnetic moment on each Ce atom; Region III corresponds to a spin density
wave that is commensurate with the lattice.\cite{Bao03}

Taken along with neutron scattering experiments on Ce$_{0.9}$La$_{0.1}$RhIn$%
_{5}$ that show no change in the magnetic structure,\cite{Bao02}
substitutions of up to 10\% of La for Ce and 10\%\ Ir for Rh do not change
the magnetic structure. This result leads one to believe that the
superconductivity in the Ir doped samples, and NFL\ behavior in the La doped
samples, evolve out of the magnetic structure of the ground state, namely an
incommensurate SDW. However, recent neutron scattering results on Ir doped
samples with $y\geq 0.30$ show the appearance of a commensurate component to
the magnetic order.\cite{Bao03}

\subsection{Long-range magnetic order}

All of the CeRh$_{1-y}$Ir$_{y}$In$_{5}$ data and Ce$_{0.97}$La$_{0.03}$RhIn$%
_{5}$ were fit using Eqs. 1-3 for $T<0.8T_{N}$. Fits were made in applied
fields of 50 kOe and 90 kOe applied along both the $a-$ and $c-$axis. A
summary of the results are displayed in Table \ref{fitdata}. 
\begin{table}[tbp]
\caption{Summary of the fitting parameters to the Ce$_{1-x}$La$_{x}$Rh$%
_{1-y} $Ir$_{y}$In$_{5}$ data. Definitions of the various coefficients are
given in the text. The units are kOe for $H$, mJ/mol Ce K$^{2}$ for $\protect%
\gamma $, mJ/mol Ce K$^{4}$ $\protect\beta _{M}$ and $\protect\beta %
_{M}^{\prime }$, K for $E_{g}/k_{B}$, mJ K/mol Ce for $A$. The number in
parentheses is the statistical uncertainty in the last digit from the least
squares fitting procedure.}\narrowtext   
\begin{tabular}{lllllllll}
& $x$ & $y$ & $H$ & $\gamma _{0}$ & $\beta _{M}$ & $\beta _{M}^{\prime }$ & $%
E_{g}/k_{B}$ & $H_{int}$ \\ 
\tableline & 0 & 0 & 0 & 50(3) & 19(1) & 510(30) & 7.0(4) & 2 \\ 
& 0 & 0 & 50($\Vert c$) & 45(2) & 24(1) & 580(30) & 7.4(4) & 56(2) \\ 
& 0 & 0 & 90($\Vert c$) & 41(2) & 29(1) & 600(30) & 7.2(4) & 97(1) \\ 
& 0 & 0 & 50($\Vert a$) & 38(2) & 19(1) & 390(30) & 5.8(4) & 54(1) \\ 
& 0 & 0 & 90($\Vert a$) & 38(2) & 6(2) & 310(30) & 4.4(2) & 94(1) \\ 
\tableline & 0 & 0.05 & 0 & 48(2) & 23(1) & 710(40) & 8.2(4) & 2 \\ 
& 0 & 0.05 & 50($\Vert c$) & 44(2) & 26(1) & 650(40) & 7.9(4) & 59(2) \\ 
& 0 & 0.05 & 90($\Vert c$) & 42(7) & 31(5) & 750(180) & 8.1(9) & 98(4) \\ 
& 0 & 0.05 & 50($\Vert a$) & 27(7) & 30(5) & 580(90) & 7.6(7) & 57(1) \\ 
& 0 & 0.05 & 90($\Vert a$) & 32(8) & 15(7) & 360(40) & 5.2(6) & 99(1) \\ 
\tableline & 0 & 0.10 & 0 & 81(4) & 15(3) & 420(60) & 6.5(4) & 2 \\ 
& 0 & 0.10 & 50($\Vert c$) & 64(3) & 27(2) & 600(50) & 8.0(4) & 57(4) \\ 
& 0 & 0.10 & 90($\Vert c$) & 56(3) & 33(2) & 680(70) & 8.1(4) & 98(1) \\ 
& 0 & 0.10 & 50($\Vert a$) & 55(3) & 25(2) & 410(40) & 6.7(4) & 57(3) \\ 
& 0 & 0.10 & 90($\Vert a$) & 38(2) & 6(2) & 310(30) & 4.4(2) & 94(2) \\ 
\tableline & 0.03 & 0 & 0 & 43(2) & 38(1) & 500(34) & 6.7(2) & 2 \\ 
& 0.03 & 0 & 50($\Vert c$) & 34(3) & 47(2) & 650(160) & 7.6(8) & 60(2) \\ 
& 0.03 & 0 & 90($\Vert c$) & 34(2) & 50(2) & 530(50) & 6.6(3) & 99(3) \\ 
& 0.03 & 0 & 50($\Vert a$) & 24(3) & 38(3) & 360(40) & 5.4(4) & 65(2) \\ 
& 0.03 & 0 & 90($\Vert a$) & 25(4) & 20(3) & 330(90) & 4.2(2) & 102(3) \\ 
&  &  &  &  &  &  &  & 
\end{tabular}%
.
\end{table}
The internal field $H_{int}$ was found by using the measured $A$ value and
using Eq. \ref{chyp}. From NQR measurements, it is known that in the absence
of an applied field the internal field at the In sites in CeRhIn$_{5}$ is of
the order 2 kOe,\cite{Curro00} which is consistent with the data in Table %
\ref{fitdata} where $H_{int}$ for the 50 and 90 kOe data is slightly higher
than the applied field. For the zero field data, a value of 2 kOe for $%
H_{int}$ gives a nearly negligible contribution to the heat capacity in our
temperature range so we have set $H_{int}=2$ kOe for $H=0$ in all of our
fits shown in Table \ref{fitdata} (this assumption has a negligible effect
on the zero field fit parameters as they do not change if $H_{int}=0$ is
used). For all values of $y$ the value of $\gamma _{0}$ is seen to decrease
as field is applied as usually seen in heavy fermion systems.\cite{Stewart84}
For $H\Vert c$ the value of $\beta _{M}$ is seen to increase as field is
applied while $E_{g}/k_{B}$ remains relatively constant. For AF\ systems
where $\beta _{M}\propto D^{-3}$ with $D$ the spin wave stiffness,\cite%
{Lees99} one would expect $D$ to decrease in an applied field that weakens
AF\ interactions leading to an increase in $\beta _{M}$. For $H\Vert a$ in
an applied field, the values of $\beta _{M}$ and $E_{g}/k_{B}$ tend to
decrease relative to the zero field value. As seen in Fig. \ref{cmfit}, the
fits to the data in this direction are for $T<T_{1}$ where the spin
structure is believed to be a SDW\ that is commensurate with the lattice. As
the zero field spin structure is an incommensurate SDW, the commensurate
state has a smaller value of $E_{g}/k_{B}$ as one would naively expect. The
small amount of La doping decreases the value of $\gamma _{0}$ as expected.
Qualitatively, applied field affects the fitting parameters for the La doped
sample in the same manner as the Ir doped samples. For $H\Vert c$ the value
of $\beta _{M}$ is seen to increase with $E_{g}/k_{B}$ remaining unchanged
within the uncertainty of the values, while for $H\Vert a$, the values of $%
\beta _{M}$ and $E_{g}/k_{B}$ both decrease as field increases.

\subsection{Short-range magnetic order}

As shown in a previous report, La doped for Ce in Ce$_{1-x}$La$_{x}$RhIn$%
_{5} $ suppresses $T_{N}$ and leads to a QCP for $x\approx 0.40$, a value
consistent with the 2D percolation threshold.\cite{Pagliuso02_2} The zero
field data for various values of $x$ ($x\leq 0.5$) are shown in Fig. \ref%
{celadata}

\begin{figure}[tbp]
\includegraphics[width=2.7in]{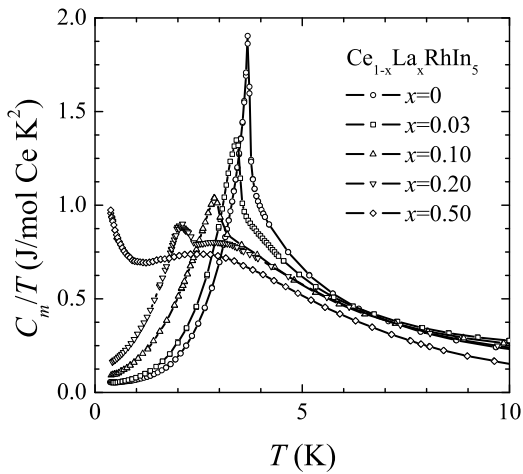}
\caption{Magnetic heat capacity $C_{m}$ divided by temperature $T$ versus $T$
measured on Ce$_{1-x}$La$_{x}$RhIn$_{5}$ in zero field.}
\label{celadata}
\end{figure}

These results are in good agreement with previous La doped \cite%
{Pagliuso02_2,Kim02} and Y doped\cite{Zapf03} reports. The value of $T_{N}$
is seen to decrease as $x$ increases indicative of a weakening of the
magnetic interactions. For $x=0.50$, the upturn at low temperatures is not
due to magnetic order but can be fit quite well for $T<0.7$ K\ to a
non-Fermi-liquid form $C/T=\gamma ^{\ast }\ln T^{\ast }/T$.\cite%
{Lohneysen94,Stewart01} In all samples, a significant portion of the entropy
($\sim R\ln 2$ as shown in Fig. \ref{entropyall}) is found above $T_{N}$.
This is consistent with neutron scattering results that show short range
magnetic correlations at temperatures on the order of $2T_{N}$.\cite{Bao02_2}
The contribution to the heat capacity from these short range correlations
can be seen as a \textquotedblleft hump\textquotedblright\ in the heat
capacity data that becomes apparent as $T_{N}$ is suppressed and is
consistent with our previous report that speculated the hump was due to
short range magnetic interactions.\cite{Pagliuso02_2}

\begin{figure}[tbp]
\includegraphics[width=2.7in]{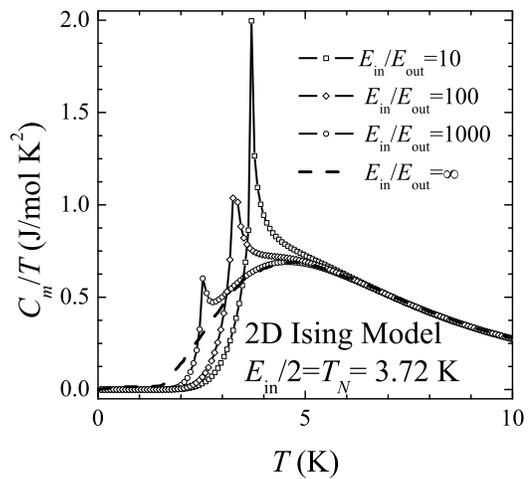}
\caption{Magnetic heat capacity $C_{m}$ divided by temperature $T$ versus $T$
as calculated for the 2D\ Ising model described in the text. $E_{in}$ and $%
E_{out}$ correspond to the magnetic interaction energies in orthogonal
directions. $E_{in}/2$ has been fixed to the value of $T_{N}=3.72$ K of
CeRhIn$_{5}$. The curves represent various values of $E_{out}.$}
\label{ising2d}
\end{figure}

To further discuss the data in terms of short range magnetic interactions,
we consider the results of McCoy and Wu for a 2D\ Ising model on a square
lattice with two magnetic interaction energies $E_{1}$ and $E_{2}$.\cite%
{Mccoy73} We identify $E_{1}$ and $E_{2}$ as $E_{in}$ and $E_{out}$ (in and
out of plane directions) that correspond to the magnetic interaction
energies along the two orthogonal directions. Where this mapping is
completely rigorous of some concern as the model only takes into account
nearest neighbor interactions on a 2D lattice, and the inclusion of both a
magnetic interaction in the c-direction along with next-nearest neighbors
would lead to multiple magnetic interaction strengths. However, to a first
approximation, it seems reasonable to qualitatively describe our data using $%
E_{in}$ as the interaction in the basal plane and $E_{out}$ as the out of
plane (interlayer) interaction. We fix $E_{in}/2=T_{N}$ and vary $E_{out}$
with the results shown in Fig. \ref{ising2d}.

The $E_{in}/E_{out}=10$ curve looks remarkably similar to the CeRhIn$_{5}$
data shown in Fig. \ref{celadata} showing a peak in $C_{m}/T$ at $T_{N}$.
Neutron scattering results on CeRhIn$_{5}$ show the magnetic correlation
lengths above $T_{N}$ are only about a factor of 2-3 different for
measurements along and perpendicular to the $c-$axis,\cite{Bao02_2} and
magnetic susceptibility measurements show a factor of 2-3 difference in the
in-plane and out-of-plane susceptibility.\cite%
{Takeuchi01,Pagliuso02_3,Christianson02,Zapf03} However, as there are more
near neighbors (4) in plane than out of plane (2) using the value of 10 for $%
E_{in}/E_{out}$ seem reasonable from a qualitative point of view. Keeping $%
E_{in}$ fixed and reducing $E_{out}$ is seen to have a dramatic effect on $%
C_{m}/T$ as $T_{N}$ moves to lower temperatures and a Schottky-like maximum
(or hump) appears. In the calculations, the heat capacity for the case $%
E_{in}/E_{out}=\infty $ shows no long range order and is identical to that
of a two-state Schottky anomaly with an energy difference of $E_{in}$
between the two levels. The evolution of the 2D Ising calculation displays
the same systematics as the data taken on Ce$_{1-x}$La$_{x}$RhIn$_{5}$ shown
in Fig. \ref{celadata}. In this scenario, as La is doped for Ce, short range
in-plane magnetic correlations remain while those along the $c-$axis are
weakened considerably by the disorder. This culminates in the observance of
no long range order for $x=0.50$.

\subsection{Non-Fermi-liquid behavior}

\begin{figure}[tbp]
\includegraphics[width=2.7in]{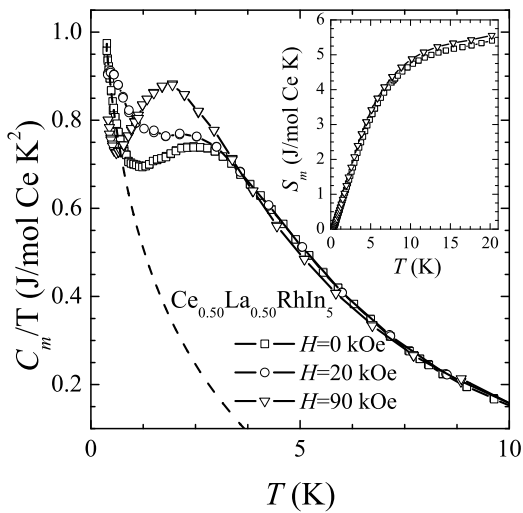}
\caption{Magnetic heat capacity $C_{m}$ divided by temperature $T$ versus $T$
measured on Ce$_{0.50}$La$_{0.50}$RhIn$_{5}$ in various applied fields. The
dashed line shows the non-Fermi liquid contribution to the heat capacity for
the zero field data as described in text. The inset shows the measured
entropy up to 20 K at 0 and 90 kOe.}
\label{cela5050}
\end{figure}

As mentioned, the heat capacity appears to display NFL\ behavior in the $%
x=0.50$ sample that is near the QCP. This is in agreement with the La doped
results of Kim \textit{et al.}\cite{Kim02} and Y doped results of Zapf 
\textit{et al}.\cite{Zapf03} who find NFL behavior near the QCP in the
magnetic susceptibility, heat capacity and resistivity (Y only) data. To
further investigate the NFL behavior in the La doped system, data in
numerous magnetic fields were collected, and some of the results are
displayed in Fig. \ref{cela5050}.

As mentioned previously, the zero field data can be fit quite well for $%
T<0.7 $ K to a NFL logarithmic dependence. The application of an applied
field moves the NFL\ feature to higher temperatures eventually appearing to
merge with the \textquotedblleft hump\textquotedblright\ centered around 2.5
K in zero field (note that the upturn at the lowest temperature for the 90
kOe applied field is due to the In nuclear term). The inset to Fig. \ref%
{cela5050} shows the measured entropy up to 20 K at 0 and 90 kOe. Note that
there is a small increase in the entropy which approaches the expected $R\ln
2$ (5.76 J/mol Ce K) as field is increased. This \textquotedblleft
missing\textquotedblright\ entropy is due to the large increase in $C_{m}/T$
at low temperatures due to the NFL\ behavior that we do not measure; if we
could measure to lower temperature, we would expect to find all of the $R\ln
2$ entropy observed for other samples (see Fig. \ref{entropyall}).

\begin{figure}[tbp]
\includegraphics[width=2.7in]{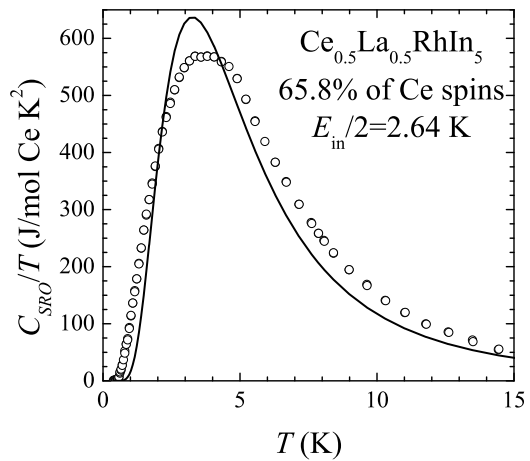}
\caption{Magnetic heat capacity due to short range magnetic order $C_{SRO}$
divided by temperature $T$ versus $T$ measured on Ce$_{0.50}$La$_{0.50}$RhIn$%
_{5}$. The solid line is a fit to the data involving 65.8\%\ of the Ce spins
as described in the text.}
\label{cehump}
\end{figure}

In heavy fermion systems, there is a natural competition between single site
and intersite interactions. This has lead to a scenario of two types of
coexisting $f$-electrons:\ a local \textquotedblleft Kondo
gas\textquotedblright\ and a global \textquotedblleft Kondo
liquid.\textquotedblright \cite{Nakatsuji02,Nakatsuji03,Curro03} In a
similar fashion, we can separate the data in Fig. \ref{cela5050} into two
components:\ a \textquotedblleft localized\textquotedblright\ term due to
short range order $C_{SRO}$ (the \textquotedblleft hump\textquotedblright )
and a NFL\ term $C_{NFL}$ that can be attributed to intersite effects. It is
important to note that the $C_{m}/T$ data in Fig. \ref{cela5050} is nearly
field independent above 5 K. Above 5 K, we would expect the NFL\
contribution to be negligible as $T^{\ast }$ in the equation $%
C_{NFL}/T=\gamma ^{\ast }\ln T^{\ast }/T$ is of the order of 5 K. Therefore,
we assume that there is a field independent term $C_{SRO}$ which we find by
subtracting off the zero field NFL\ contribution. The resulting $C_{SRO}$ is
displayed in Fig. \ref{cehump}.

The data can be fit reasonably well by the 2D Ising model with a value of $%
E_{in}/2=2.64$ K. and only around 2/3 of the Ce spins being involved. These
numbers are similar to those obtained on Y doped samples.\cite{Zapf03}
However Zapf \textit{et al}.\cite{Zapf03} interpret their data in terms of
crystal fields that also display Schottky-like behavior. As mentioned
previously, the heat capacity in the 2D\ Ising model for the case $%
E_{in}/E_{out}=\infty $ is identical to that of a two-state Schottky anomaly.%
\cite{Mccoy73} The fact that the data is broader than the fit is likely due
to disorder. McCoy and Wu have indeed shown that disorder broadens measured
features in heat capacity measurements.\cite{Mccoy73} The value of $E_{in}/2$
is less than the value of $T_{N}$ for the undoped sample that is used in
Fig. \ref{ising2d}. As already discussed, doping non magnetic La atoms
should not only reduce $E_{out}$ as was done in Fig. \ref{ising2d}, it
should also reduce $E_{in}$ as we observe. The short range order scenario
also gives a natural explanation of the absence of magnetic field effects on
the $C_{SRO}$ data because the spins align in the basal plane and the
magnetic field is applied perpendicular to the spins. The magnetic field
would be expected to have an effect on $E_{out}$ while leaving $E_{in}$
unchanged. Since $E_{out}\ll E_{in}$ the ratio of $E_{in}/E_{out}$ will be
quite large and the heat capacity data is very insensitive to changes in $%
E_{out}$ as long as the ratio is large, and $C_{SRO}$ appears Schottky-like
in nature peaking at the same temperature since $E_{in}$ is not changing.
The $C_{SRO}$ data is also similar to that seen by others near the QCP for
La doping,\cite{Kim02} Y\ doping,\cite{Zapf03} and applied pressure.\cite%
{Fisher02}

The case of high pressure is of particular interest, because the
\textquotedblleft hump\textquotedblright\ like feature is seen in the
undoped stoichiometry near the QCP\ when pressure is applied.\cite{Fisher02}
Though Fisher \textit{et al}. attribute the \textquotedblleft
hump\textquotedblright\ to the Kondo effect,\cite{Fisher02} the feature is
much too narrow to be fit by a spin 1/2 Kondo impurity model.\cite{Rajan83}
In fact the feature due to superconductivity is seen below a maximum in $C$
that is field independent as is our $C_{SRO}$ data. In this interpretation,
the long range 3D magnetic order is nearly destroyed while short range 2D\
magnetic correlations are still present near the QCP. This could easily be
envisioned by the magnetic correlation length becoming smaller along the
c-axis than the nearest neighbor Ce-Ce separation which is much greater than
the basal plan Ce-Ce distances leading to very large values of $%
E_{in}/E_{out}$. For the 16.5 kbar data of Fisher \textit{et al}. that is
very near the QCP, the data is fit fairly well assuming that 65\% of the Ce
spins are involved in the magnetic heat capacity with an energy of $%
E_{in}/2=2.45$ K. These numbers are remarkably similar to the Ce$_{0.5}$La$%
_{0.5}$RhIn$_{5}$ data that is near the QCP. The small feature near 2.5 K in
the pressure data could be explained by AF\ order with the fit parameters
listed. This would mean that the superconducting transition would be at even
lower temperatures (perhaps the very small looking feature in the data at 1
K).\cite{Fisher02} Recent NQR\ results at 16 kbar are interpreted in terms
of microscopic regions of AF order below around 2.8 K and AF\ and
superconducting states below 1.3 K.\cite{Mito03} This scenario is in
excellent agreement with the above analysis. The application of higher
pressure would be expected to increase the anisotropy of the magnetic
interactions while reducing the magnitude of $E_{in}/2$. This is exactly
what is found at 19 kbar where the data is fit reasonably well by $%
E_{in}/2=1.88$ K with $E_{in}/E_{out}=\infty .$ Unlike the crystal field
interpretation,\cite{Zapf03} the universality of the \textquotedblleft
hump\textquotedblright\ like feature near the QCP\ in both pressure and
doping experiments, the agreement with the 2D Ising calculations, and the
fact that the crystal field levels are found not to change with La doping in
CeCoIn$_{5}$ compounds\cite{Nakatsuji02} lead us to confidently conclude
that the data we have labelled $C_{SRO}$ is indeed due to short range
magnetic order.

The remaining $\sim 34\%$ of the entropy for the $x=0.50$ sample is found in
a NFL\ $\gamma ^{\ast }\ln T^{\ast }/T$ term. After the nuclear In term is
estimated using Eq. \ref{chyp} with $H_{int}=H$ and subtracted along with
the field independent $C_{SRO}/T$ contribution, the remaining data is $%
C_{NFL}/T$. The $C_{NFL}/T$ data is plotted for various fields in Fig. \ref%
{cenfl}.

\begin{figure}[tbp]
\includegraphics[width=2.7in]{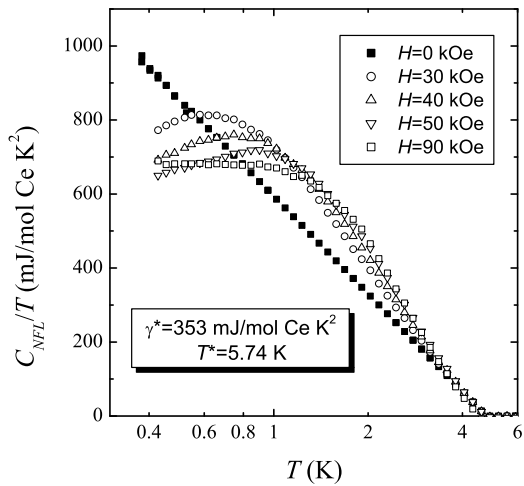}
\caption{Measured non-Fermi-liquid heat capacity $C_{NFL}$ divided by
temperature $T$ versus $T $ measured on Ce$_{0.50}$La$_{0.50}$RhIn$_{5}$ in
various applied fields. The zero field data is given by $\protect\gamma %
^{\ast }\ln T^{\ast }/T$ with $\protect\gamma ^{\ast }=353$ mJ/mol Ce K$^{2}$
and $T^{\ast }=5.7$ K.}
\label{cenfl}
\end{figure}

As field increases, the low temperature data appears to go to a nearly
constant Fermi-liquid like value. This field dependent behavior is
remarkably similar to that seen in one of the prototypical NFL\ system CeCu$%
_{5.9}$Au$_{0.1},$\cite{Lohneysen94} and also to recent results on CeCoIn$%
_{5}$.\cite{Bianchi03} These findings lead to the conclusion that the $4f$%
-electrons in Ce$_{0.5}$La$_{0.5}$RhIn$_{5}$ display both short range (short
range magnetic order) and long range (non-Fermi liquid) behavior consistent
with an evolving picture of coexisting short and long range correlations in
the Kondo lattice.\cite{Nakatsuji02,Nakatsuji03,Curro03}

\section{Conclusions}

In conclusion, we have measured the heat capacity on both the CeRh$_{1-y}$Ir$%
_{y}$In$_{5}$ $(y\leq 0.10)$ and Ce$_{1-x}$La$_{x}$RhIn$_{5}$ $(x\leq 0.50)$
systems. Field induced transitions when the magnetic field is applied along
the $a$ direction for CeRh$_{0.90}$Ir$_{0.10}$In$_{5}$ are very similar to
those observed in CeRhIn$_{5}$ suggestive that Ir doping up to 10\% does not
change the AF\ order from the measured incommensurate spin density wave in
the undoped sample.\cite{Bao00} La doping suppresses magnetic order with the 
$x=0.50$ sample showing no long range AF\ order. The La doped data shows
excellent agreement to calculation on a 2D\ square Ising lattice with La
doping weakening the out of plane magnetic interactions. In Ce$_{0.50}$La$%
_{0.50}$RhIn$_{5}$, a coexistence of short range magnetic order and
non-Fermi-Liquid behavior is observed that is remarkably similar in nature
to the pressure induced superconductivity at a QCP\ in undoped CeRhIn$_{5}$.%
\cite{Fisher02} This \textquotedblleft dual\textquotedblright\ nature of the 
$4f$-electrons in Ce-based and other heavy fermion systems shows the
importance of including the effect of the lattice when studying these
systems.

\begin{acknowledgements}
Work at UNLV is supported by DOE EPSCoR-State/National Laboratory Partnership
Award DE-FG02-00ER45835 and DOE Cooperative Agreement DE-FC08-98NV1341 Work at
LANL is performed under the auspices of the U.S. Department of Energy.
\end{acknowledgements}

$^*$ Present address: Instituto de Física " Gleb Wataghin " - UNICAMP, 13083-970, Campinas, Brazil

\newif\ifabfull\abfulltrue


\begin{thebibliography}{47}
\bibitem[{fir()}]{firstord}
\bibinfo{note}{The software used to measure the heat capacity uses average
  values to fit the thermal relaxation data. From the raw traces, the
  transitions which we have assigned as first order are clearly first order in
  nature. Experience tells us the averaging method underestimates the value of
  the heat capacity near the peak of the first order transition by a factor of
  four or five.}
\end{thebibliography}
\end{document}